\documentclass[aps,pra,twocolumn]{revtex4-2}

\usepackage{amsmath}
\usepackage{amssymb}

\usepackage{setspace}
\usepackage{bm}
\usepackage{graphicx}
\numberwithin{equation}{section}
\usepackage{hyperref}
\usepackage[utf8]{inputenc} 

\usepackage{braket} 

\newcommand{\udubee}{Department of Electrical and Computer Engineering, University of Washington, Seattle, WA 98195, USA}
\newcommand{\udubphys}{Department of Physics, University of Washington, Seattle, WA 98195, USA}

\begin{document}
\title{Image amplification in a self-imaging degenerate optical cavity}
\author{Albert Ryou$^1$, Shane Colburn$^1$, Arka Majumdar$^{1,2}$}
\affiliation{
$^1$ \udubee \\
$^2$ \udubphys}
\date{\today}

\begin{abstract}

Image-amplifying optical cavities offer an intriguing platform for nonlinear optical processing of two-dimensional signals, with potential applications in optical information processing and optically implemented artificial neural networks. Here, we analyze the performance of a self-imaging degenerate cavity via numerical simulations and show the existence of a cavity size-dependent minimum spread in the transverse mode resonances. This non-degeneracy, in turn, leads to an inherent trade-off between the amplification and the fidelity of the intracavity image as the cavity finesse changes. Our results point to a promising path to nonlinear image processing in a low-power, small-form-factor optical cavity.

\end{abstract}

\maketitle


\section{Introduction}

Nonlinear processing of two-dimensional optical signals, or images, holds great potential for many research fields, including optical information processing \cite{Cotter1523}, parametric amplification \cite{gigan2006continuous}, and optical artificial neural networks \cite{shen2017deep, lin2018all, colburn2018optical}. Generating optical nonlinearities, however, typically requires high-intensity pulsed lasers, limiting the activity to only laboratory settings. An alternative method is to employ an optical cavity to resonantly enhance the intracavity field of a continuous-wave laser. While cavities have long been used for nonlinear optics \cite{kozlovsky1988efficient, boyd1992nonlinear, fryett2016silicon}, they have generally been operated in single mode, amplifying one specific optical frequency and suppressing the rest of the spectrum. Because the modes also have distinct spatial patterns, a typical cavity acts as both a spectral and a spatial filter. An attempt to inject a monochromatic image into such a cavity will thus result in its decomposition, with only the resonance-matching spatial component being transmitted.

To amplify an image as a whole, the transverse modes that comprise the image must be spectrally degenerate. A class of cavities known as degenerate cavities exhibit multiple modes that satisfy the same resonance condition. For a special subset, called self-imaging cavities, all transverse modes are completely degenerate \cite{arnaud1969degenerate}. These concepts have been the basis for several works: Gigan et al. used a hemiconfocal cavity, which exhibits four degenerate families of modes, to demonstrate pattern formation \cite{gigan2005image}; and Chalopin et al. used a self-imaging cavity to frequency double an image \cite{chalopin2010frequency}.

\begin{figure}
\includegraphics[width=83mm]{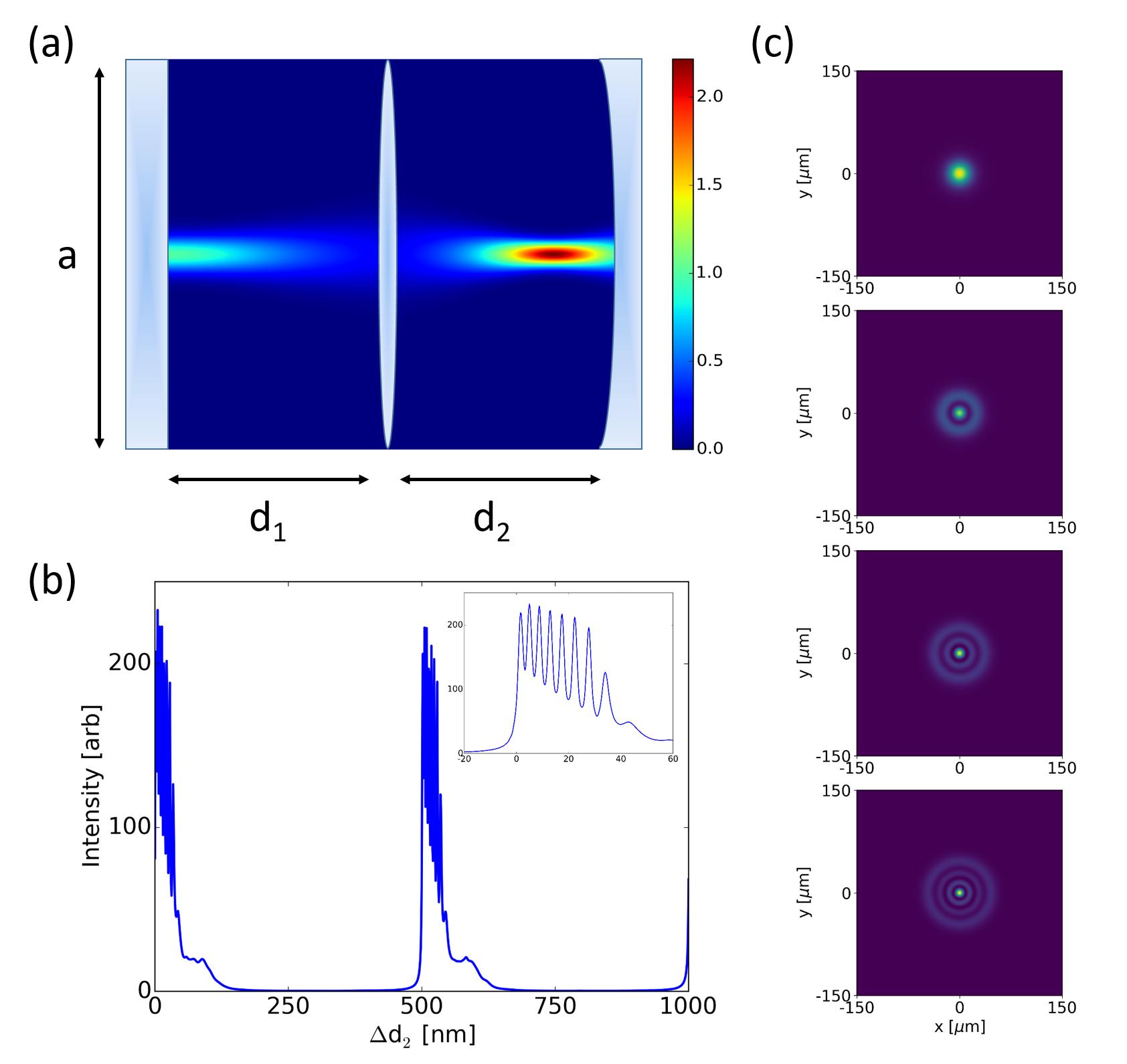}
\caption{\label{Figure:cavity_cartoon} (Color online) \textbf{Cavity geometry and mode spectrum}. (a) Schematic diagram of the cavity. The cavity consists of a flat mirror, a thin lens with focal length $f$, and a curved mirror with radius of curvature $R$, separated by $d_1$ and $d_2$. The optical elements have a transverse extent defined by the aperture diameter $a$. (b) Mode spectrum of a cavity with $f = R = d_1/2 = 1$ mm, $a = 300$ $\mu$m, and $d_2 = d_1 + \Delta d_2$. The operating wavelength $\lambda$ is 1 $\mu$m. The amplitude reflection coefficients of the mirrors are set to 0.96, while the lens exhibits unity transmission; (inset) zoomed-in plot of the first set of peaks. The profile of the fundamental mode (first peak) has been overlaid in (a), where the color represents the field intensity. (c) The transverse profiles of the first four peaks are well-approximated by the Laguerre-Gauss (LG) modes.} 
\end{figure}

\begin{figure}
\includegraphics[width=83mm]{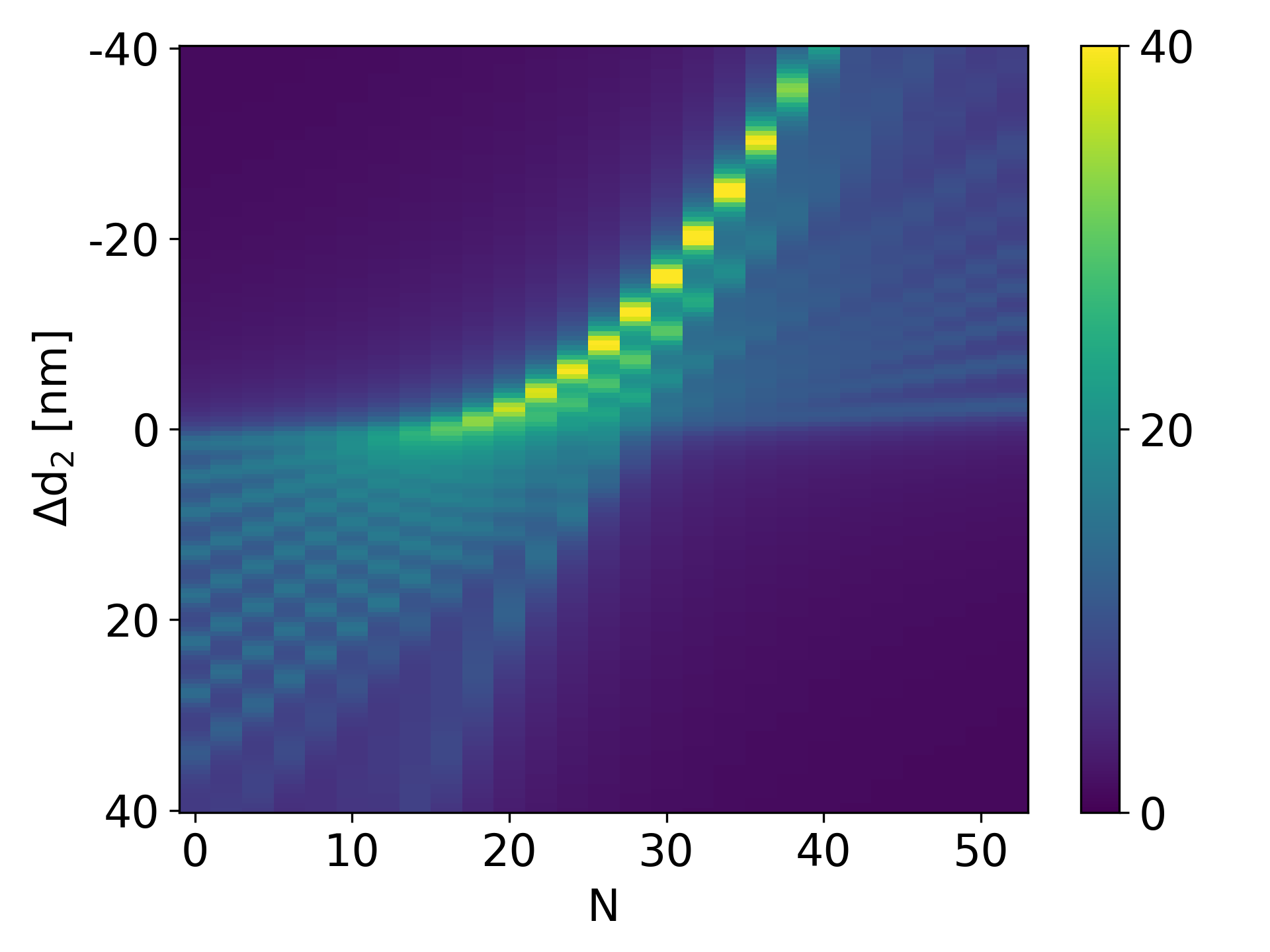}
\caption{\label{Figure:cavity_cartoon} (Color online) \textbf{Mode spectrum vs. cavity size}. Plot of the cavity mode spectrum as a function of the change in $d_2 = d_1 + \Delta d_2 - N \lambda/2$, with $\Delta d_2$ on the y-axis and axial mode shift number $N$  on the x-axis. As $N$ increases from zero, the modes first become increasingly closely-packed before reaching the minimum spread, after which they begin to spread out again. The colorbar represents the square root of the field intensity.} 
\end{figure}


Recent advances in nanophotonics have made it timely and relevant to revisit the use of the self-imaging degenerate cavity in the context of miniaturized optics. On the other hand, when optical elements shrink in size, their functionalities can change in unanticipated ways \cite{ghosh2011miniaturized, liu2016fundamental}. In particular, the paraxial approximation, which forms the bedrock of conventional laser resonator theory, can no longer be taken for granted. The effect of the approximation's breakdown on the mode degeneracy, and subsequently on the amplification and the fidelity of the intracavity image, warrants careful investigation.

In this paper, we analyze a self-imaging degenerate cavity and explore the consequence of the non-degeneracy of the transverse modes. Using Fourier optics to model light propagation, we demonstrate the formation of individual modes as well as the coherent buildup of an intracavity image. We then study the relationships among image amplification, fidelity, and cavity size. Finally, we investigate the trade-off between amplification and fidelity resulting from tuning the cavity finesse. 
	

\section{Cavity Simulation}

The cavity, shown in Fig. 1a, is composed of three optical elements: a flat mirror, a thin converging lens with focal length $f$, and a curved mirror with radius of curvature $R$. The distance between the flat mirror and the lens is $d_1$, and the distance between the lens and the curved mirror is $d_2$. All the optical elements share the same transverse extent denoted by the aperture diameter $a$. The two mirrors are partially reflective with the amplitude reflection coefficient $r$, while the lens exhibits unity transmission. 

To simulate light propagation inside the cavity, we use an iterative approach based on the Fox and Li method \cite{fox1961resonant} (see Appendix A). The modes of the cavity are found by injecting a plane wave from the side of the flat mirror. Each round trip yields a modified field profile, and the total intensity, given by the absolute square of the sum of the individual fields after each round trip, reaches a steady state after a sufficient number of round trips.

Computing the cavity mode spectrum requires repeating the process and calculating a series of total intensities while tuning one of its parameters, typically either the wavelength of the incident light or the length of the cavity. We chose to tune the cavity length by sweeping $d_2$, the distance between the lens and the curved mirror, while leaving all other parameters fixed. Figure 1b shows a plot of the mode spectrum of a cavity with $f = R = d_1/2 = 1$ mm, $a = 300$ $\mu$m, and $d_2 = d_1 + \Delta d_2$, where $\Delta d_2$ runs from zero to $\lambda = 1$ $\mu$m. As expected, the spectrum exhibits two sets of sharp, closely-packed peaks, one for each axial mode of the cavity. The inset shows a zoomed-in view of the first set. Figure 1c shows the intensity profiles of the first four peaks in the set, which resemble the lowest-order Laguerre-Gauss (LG) modes with $l = 0$ and $p = 0, 1, 2$, and $3$ \cite{siegman1986lasers}.


\begin{figure}
\includegraphics[width=83mm]{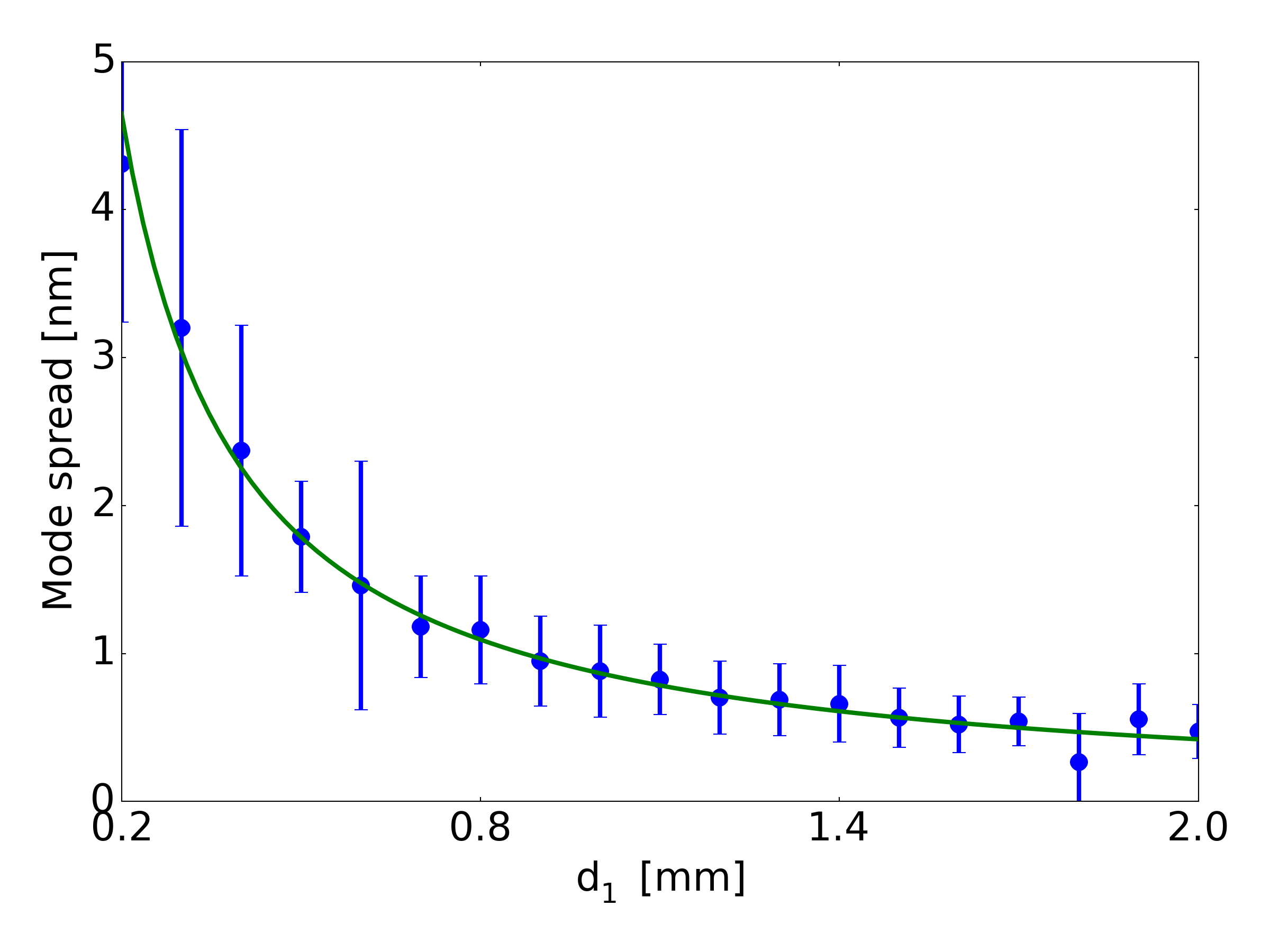}
\caption{\label{Figure:cavity_cartoon} (Color online) \textbf{Minimum mode spread vs. cavity size}. Plot of the minimum spread as a function of $d_1$, which parametrizes the cavity length. The aperture diameter of the cavity is set to $\sqrt{50} w$, where $w$ is the largest beam width of the fundamental mode inside the cavity. The spread is smaller for larger $d_1$, indicating that the cavity becomes more degenerate as its dimensions grow. The curve is a guide to the eyes only. } 
\end{figure}


\section{Analysis of the mode degeneracy}

For infinite-aperture paraxial cavities, it has been shown that the Gouy phase that is responsible for the distinct frequencies of the transverse modes becomes zero for several cavity configurations, leading to the completely degenerate spectrum \cite{arnaud1969degenerate} (see Appendix B). For our cavity, the degeneracy is predicted to occur when $f = R = d_1/2 = d_2/2$. As Fig. 1b shows, however, the resulting modes, although closely-packed, are not degenerate. We attribute this discrepancy to the combination of several factors, including the focal shift and the non-paraxial nature of the intracavity light, as elaborated below. A contribution from the numerical effect stemming from the cavity's finite aperture is described in Appendix C. 

  

\begin{figure*}
\includegraphics[width=163mm]{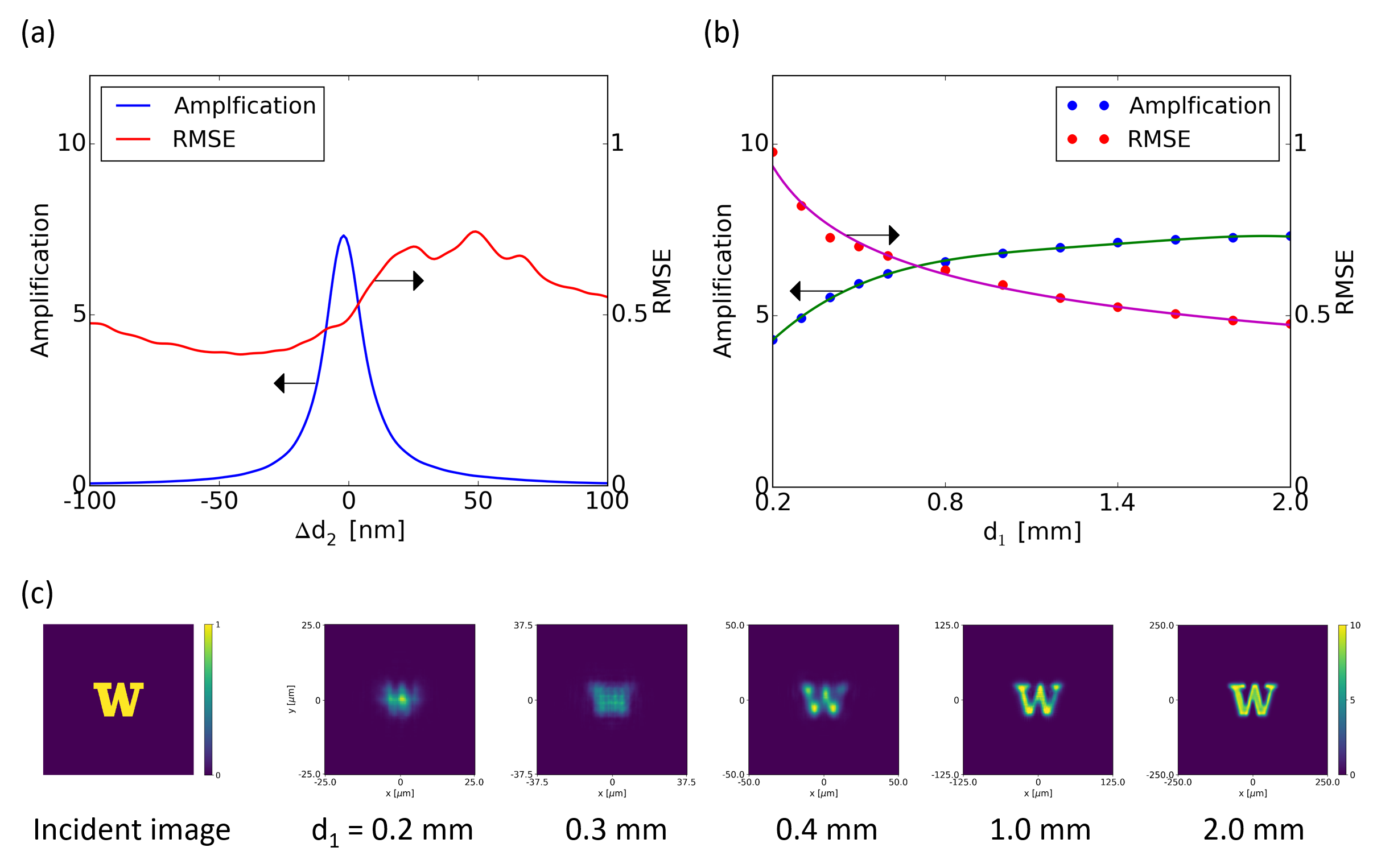}
\caption{\label{Figure:cavity_cartoon} (Color online) \textbf{Injecting an image}. (a) Plot of the intracavity field intensity amplification (blue) and RMSE (red) vs. $\Delta d_2$. The amplification is the highest where the cavity modes occur. (b) Amplification (blue) and RMSE (red) vs. cavity size, parametrized by $d_1$. As can be predicted from the result of Fig. 3, as the cavity size decreases, the image becomes both less intense and less similar to the incident image. The error bars are smaller than the points. The curves are guides to the eyes only. (c) Intensity profiles of the incident image, along with the intracavity image for $d_1 = 0.2, 0.3, 0.4, 1.0$, and $2.0$ mm.}
\end{figure*}


It is well-known that a collimated beam incident on an ideal lens focuses slightly inside the lens' geometrical focus $f$ \cite{self1983focusing}. The deviation, known as the focal shift, is given by $\Delta f = z_R^2/f$, where the Rayleigh range $z_R = \pi w^2/\lambda$, and $w$ is the $1/e^2$ intensity radius of the fundamental cavity mode at a focus \cite{siegman1986lasers}. For typical focused beams, $z_R \ll f$ such that $\Delta f$ is negligible; for our degenerate cavity, on the other hand, both $f$ and $z_R$ near the flat mirror are on the order of a millimeter, resulting in non-zero $\Delta f$. This suggests then that we may compensate for the focal shift and reduce the spread among the modes by making the cavity shorter. To do this, we shift $d_2$ away from the predicted degeneracy condition by setting $d_2 = d_1 + \Delta d_2 - N \lambda/2$, where $\Delta d_2$ remains the sub-wavelength detuning of the curved mirror as before, and $N$ is an integer denoting the number of axial modes by which we shift $d_2$. Figure 2 shows a plot of the mode spectrum as a function of $\Delta d_2$ (y-axis) and $N$ (x-axis) for a cavity with $f = R = d_1/2 = 1$ mm, $r = 0.993$, and $a = 300$ $\mu$m. As $N$ increases, the modes become increasingly closely-packed. Eventually, they reach a point of minimum spread before fanning out again.
  

We attribute the remaining minimum spread to the breakdown of the paraxial approximation. Gaussian optics and the conventional laser resonator theory that employs the Gouy phase are built on the paraxial approximation that an optical wavefront propagates only at a small angle relative to the optical axis \cite{siegman1986lasers}. The degree to which the approximation holds thus depends on the modes' beam width $w$, with first-order corrections for the non-paraxial field given by $(\lambda/w)^2$ \cite{agrawal1979gaussian}. For our cavity with $\lambda = 1$ $\mu$m and $w < 10 $ $\mu$m near the curved mirror, it is possible that the non-paraxial nature of the degenerate cavity contributes to the spread among the modes.

Finally, we study how this minimum spread, which has a direct consequence on the intracavity image quality, depends on the size of the cavity. Intuitively, as the cavity shrinks and $w$ becomes smaller, the intracavity field should become more non-paraxial, causing the modes to spread further. Figure 3 shows the minimal spread, now defined to be the smallest range achievable among the four lowest LG modes by tuning $\Delta d_2$, versus the cavity size, which is parametrized by $d_1$. While sweeping $d_1$, we set $f = R = d_1/2$, $a = \sqrt{50}w$, where $w$ is the beam width at the plane of the lens, and $N$ is chosen to produce the smallest spread for each simulation. As can be seen in Fig. 3, the minimum spread has an inverse relationship with the cavity length, the effect of which we explore further in the following section.


\section{Image amplification and fidelity vs cavity size}

The buildup of an image inside a degenerate cavity is the result of the coherent superposition of the individual modes. The fact that the modes occur at different cavity lengths, however, means that any fixed $d_2$ will lead to unequal enhancement of the modes, in turn leading to image distortion. A realistic image may consist of hundreds of cavity modes, and it is difficult to predict the amount of distortion in the image as a function of the spread in the modes, especially since the act of perceiving image quality is largely psychological \cite{phillips2018camera}. For this reason, we chose to quantify the quality of the intracavity image with the intensity amplification and the root-mean-square-error (RMSE), both calculated by averaging over all pixels within the boundary of the original incident image that we shine on the cavity. A large RMSE denotes low image fidelity and vice versa.


Intuitively, a cavity whose modes are more closely-packed should form an intracavity image with higher amplification and fidelity. To test this idea, for a given cavity size, we perform the cavity simulation, but instead of exciting with a plane wave, we inject an image of a ``W'' with an uniform intensity. The image is placed in the middle of the cavity, and its lateral dimension is set to a third of the aperture diameter $a$ (see Appendix D for the effect of different incident image sizes). Figure 4a shows the amplification (blue) and the RMSE (red) of the intracavity image as we tune $d_2$ near the minimal-spread point found in the previous section. The cavity parameters are $f = R = d_1/2 = 1$ mm, $d_2 = d_1 + \Delta d_2 - N \lambda/2$, $N = 16$, $a = 250$ $\mu$m, and $r = 0.95$. As expected, the amplification is the highest near where the modes occur.

We then analyze the performance of the cavity as it shrinks. We again parametrize the cavity length by $d_1$, while keeping $f=R=d_1/2$. The aperture diameter is given by $a = d_1/4$. Thus, both the transverse and the axial dimensions of the cavity maintain their aspect ratio as the cavity size becomes reduced. The size of the incident image also maintains a fixed ratio relative to $a$. For each $d_1$, we identify $N$ and $\Delta d_2$ that yield the highest amplification and record the values of the amplification and the RMSE. Figure 4b shows the result of varying $d_1$ from 0.2 mm to 2 mm. As the cavity shrinks, the amplification decreases and the RMSE increases, which are expected from the cavity size-dependent spread in the modes observed in Fig. 3. Figure 4c shows the intensity profile of the incident image, along with the intracavity intensity for cavities with different $d_1$. For $d_1 = 0.2$ mm, the intracavity image is virtually unrecognizable; as $d_1$ approaches $2.0$ mm, it becomes progressively sharper and more similar to the incident image.




\section{Image amplification and fidelity vs cavity finesse}


\begin{figure}
\includegraphics[width=83mm]{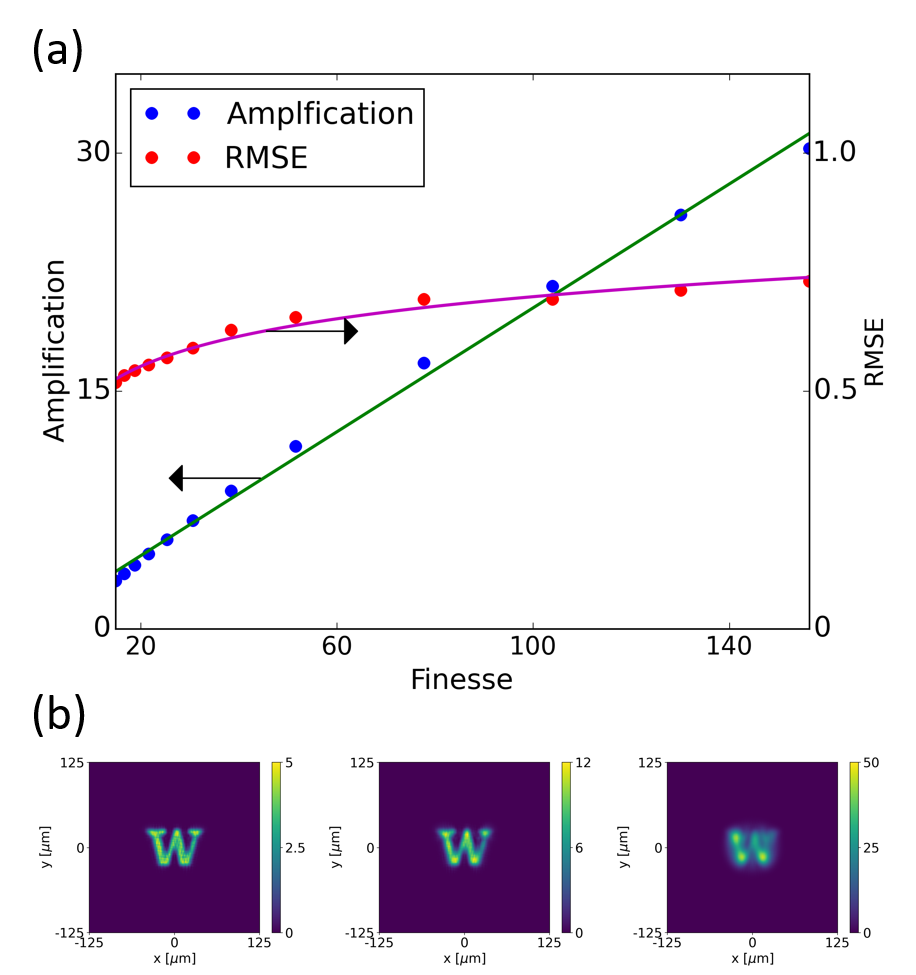}
\caption{\label{Figure:cavity_cartoon} (Color online) \textbf{Intracavity image vs cavity finesse}. (a) Plot of the amplification (blue) and RMSE (red) as a function of the cavity finesse. For high mirror reflectivity and high finesse, the modes become more amplified, but the disparity among the individual modes' amplifications grows. This, in turn, increases the RMSE and distorts the image. On the other hand, for low mirror reflectivity and low finesse, the image is less amplified but appears more similar to the incident image. The error bars are smaller than the points. The curves are guides to the eyes only. (b) Transverse profiles of the intracavity field intensity for $F \approx 15, 31$, and $160$.} 
\end{figure}


While the image amplification and fidelity are largely determined by the cavity size-dependent spread in the modes, it is possible to increase the effective degeneracy by modifying the cavity finesse. The linewidth $\Delta l$ of the modes, seen in Fig. 1b, is determined by $\Delta l = \lambda/F$, where the finesse $F$ is a function of the amplitude reflection coefficient $r$ of the mirrors via $F = \pi r/(1-r^2)$. By decreasing $r$ and therefore $F$, we can increase $\Delta l$, and consequently, the ratio of the amplifications for any pair of modes becomes closer to unity. Thus, although the mode locations remain unchanged, different modes undergo a more equal enhancement, yielding an intracavity image with higher fidelity as a result. The cost of this effective degeneracy is the reduction in the image amplification.

Figure 5a shows the amplification (blue) and the RMSE (red) as a function of the finesse. Both rise as $F$ increases. The intensity profiles shown in Fig. 5b confirm our expectations. For $F \approx 15$, the intracavity image sharply resembles the incident image but with low amplification ($\sim 5$); for $F \approx 160$, the amplification is high ($\sim 50$), but the image appears highly distorted. Thus, depending on the incident image and its modal composition, one can exchange the intracavity amplification, and hence the efficiency of the nonlinear optical processes, for additional image fidelity. The required level of fidelity in turn may be dictated by specific applications, for instance, the required image classification accuracy for a desired deep learning task.

\section{Conclusion}

We have simulated a self-imaging degenerate cavity and analyzed the effect of its mode degeneracy on the intracavity image. The results show the critical role that the non-paraxial nature of light plays for image-enhancing cavities even for seemingly macroscopic dimensions on the order of a millimeter. The trade-off among the intracavity image amplification, fidelity, and cavity size must be carefully considered when employing a self-imaging cavity for specific applications, such as nonlinear image processing with $\chi_2$, $\chi_3$, self-electro-optic, or two-dimensional materials \cite{nozaki2010sub, majumdar2014cavity, fryett2015cavity, liu2019van} and designing robust arbitrary optical potentials for experiments in cavity quantum electrodynamics \cite{jaksch1998cold, jia2018strongly}.

Building such a cavity with conventional optical elements may be experimentally challenging, due to the required alignment accuracy on the order of a few nanometers, as can be seen from the amplification peak in Fig. 4a. Multiple precisely-aligned optical elements, on the other hand, can be fabricated by using stacked sub-wavelength diffractive elements such as metasurfaces \cite{yu2014flat, zhan2017metasurface, colburn2018metasurface}, which enable drift- and misalignment-resistant monolithic designs with an exceptionally small footprint. For use in neural networks, the reflectivities of the distributed Bragg reflectors may be chosen to increase either the amplification or the fidelity at the expense of the other, according to the available training and overall computational resources \cite{colburn2018optical}.

\section{Acknowledgements}

This work was supported by the UW Royalty Research Fund and Air Force grant FA9550-18-1-0104. A.R. acknowledges support from the IC Postdoctoral Research Fellowship and the Mistletoe Research Fellowship. A.M. acknowledges support from the Alfred P. Sloan Research Fellowship. 





\bibliography{TheBib}{}

\begin{thebibliography}{30}%
\makeatletter
\providecommand \@ifxundefined [1]{%
 \@ifx{#1\undefined}
}%
\providecommand \@ifnum [1]{%
 \ifnum #1\expandafter \@firstoftwo
 \else \expandafter \@secondoftwo
 \fi
}%
\providecommand \@ifx [1]{%
 \ifx #1\expandafter \@firstoftwo
 \else \expandafter \@secondoftwo
 \fi
}%
\providecommand \natexlab [1]{#1}%
\providecommand \enquote  [1]{``#1''}%
\providecommand \bibnamefont  [1]{#1}%
\providecommand \bibfnamefont [1]{#1}%
\providecommand \citenamefont [1]{#1}%
\providecommand \href@noop [0]{\@secondoftwo}%
\providecommand \href [0]{\begingroup \@sanitize@url \@href}%
\providecommand \@href[1]{\@@startlink{#1}\@@href}%
\providecommand \@@href[1]{\endgroup#1\@@endlink}%
\providecommand \@sanitize@url [0]{\catcode `\\12\catcode `\$12\catcode
  `\&12\catcode `\#12\catcode `\^12\catcode `\_12\catcode `\%12\relax}%
\providecommand \@@startlink[1]{}%
\providecommand \@@endlink[0]{}%
\providecommand \url  [0]{\begingroup\@sanitize@url \@url }%
\providecommand \@url [1]{\endgroup\@href {#1}{\urlprefix }}%
\providecommand \urlprefix  [0]{URL }%
\providecommand \Eprint [0]{\href }%
\providecommand \doibase [0]{https://doi.org/}%
\providecommand \selectlanguage [0]{\@gobble}%
\providecommand \bibinfo  [0]{\@secondoftwo}%
\providecommand \bibfield  [0]{\@secondoftwo}%
\providecommand \translation [1]{[#1]}%
\providecommand \BibitemOpen [0]{}%
\providecommand \bibitemStop [0]{}%
\providecommand \bibitemNoStop [0]{.\EOS\space}%
\providecommand \EOS [0]{\spacefactor3000\relax}%
\providecommand \BibitemShut  [1]{\csname bibitem#1\endcsname}%
\let\auto@bib@innerbib\@empty
\bibitem [{\citenamefont {Cotter}\ \emph {et~al.}(1999)\citenamefont {Cotter},
  \citenamefont {Manning}, \citenamefont {Blow}, \citenamefont {Ellis},
  \citenamefont {Kelly}, \citenamefont {Nesset}, \citenamefont {Phillips},
  \citenamefont {Poustie},\ and\ \citenamefont {Rogers}}]{Cotter1523}%
  \BibitemOpen
  \bibfield  {author} {\bibinfo {author} {\bibfnamefont {D.}~\bibnamefont
  {Cotter}}, \bibinfo {author} {\bibfnamefont {R.~J.}\ \bibnamefont {Manning}},
  \bibinfo {author} {\bibfnamefont {K.~J.}\ \bibnamefont {Blow}}, \bibinfo
  {author} {\bibfnamefont {A.~D.}\ \bibnamefont {Ellis}}, \bibinfo {author}
  {\bibfnamefont {A.~E.}\ \bibnamefont {Kelly}}, \bibinfo {author}
  {\bibfnamefont {D.}~\bibnamefont {Nesset}}, \bibinfo {author} {\bibfnamefont
  {I.~D.}\ \bibnamefont {Phillips}}, \bibinfo {author} {\bibfnamefont {A.~J.}\
  \bibnamefont {Poustie}},\ and\ \bibinfo {author} {\bibfnamefont {D.~C.}\
  \bibnamefont {Rogers}},\ }\bibfield  {title} {\bibinfo {title} {Nonlinear
  optics for high-speed digital information processing},\ }\href
  {https://doi.org/10.1126/science.286.5444.1523} {\bibfield  {journal}
  {\bibinfo  {journal} {Science}\ }\textbf {\bibinfo {volume} {286}},\ \bibinfo
  {pages} {1523} (\bibinfo {year} {1999})}\BibitemShut {NoStop}%
\bibitem [{\citenamefont {Gigan}\ \emph {et~al.}(2006)\citenamefont {Gigan},
  \citenamefont {Lopez}, \citenamefont {Delaubert}, \citenamefont {Treps},
  \citenamefont {Fabre},\ and\ \citenamefont
  {Ma{\^\i}tre}}]{gigan2006continuous}%
  \BibitemOpen
  \bibfield  {author} {\bibinfo {author} {\bibfnamefont {S.}~\bibnamefont
  {Gigan}}, \bibinfo {author} {\bibfnamefont {L.}~\bibnamefont {Lopez}},
  \bibinfo {author} {\bibfnamefont {V.}~\bibnamefont {Delaubert}}, \bibinfo
  {author} {\bibfnamefont {N.}~\bibnamefont {Treps}}, \bibinfo {author}
  {\bibfnamefont {C.}~\bibnamefont {Fabre}},\ and\ \bibinfo {author}
  {\bibfnamefont {A.}~\bibnamefont {Ma{\^\i}tre}},\ }\bibfield  {title}
  {\bibinfo {title} {Continuous-wave phase-sensitive parametric image
  amplification},\ }\href {https://doi.org/10.1080/09500340500331364}
  {\bibfield  {journal} {\bibinfo  {journal} {J. Mod. Opt.}\ }\textbf {\bibinfo
  {volume} {53}},\ \bibinfo {pages} {809} (\bibinfo {year} {2006})}\BibitemShut
  {NoStop}%
\bibitem [{\citenamefont {Shen}\ \emph {et~al.}(2017)\citenamefont {Shen},
  \citenamefont {Harris}, \citenamefont {Skirlo}, \citenamefont {Prabhu},
  \citenamefont {Baehr-Jones}, \citenamefont {Hochberg}, \citenamefont {Sun},
  \citenamefont {Zhao}, \citenamefont {Larochelle}, \citenamefont {Englund},\
  and\ \citenamefont {Soljacic}}]{shen2017deep}%
  \BibitemOpen
  \bibfield  {author} {\bibinfo {author} {\bibfnamefont {Y.}~\bibnamefont
  {Shen}}, \bibinfo {author} {\bibfnamefont {N.~C.}\ \bibnamefont {Harris}},
  \bibinfo {author} {\bibfnamefont {S.}~\bibnamefont {Skirlo}}, \bibinfo
  {author} {\bibfnamefont {M.}~\bibnamefont {Prabhu}}, \bibinfo {author}
  {\bibfnamefont {T.}~\bibnamefont {Baehr-Jones}}, \bibinfo {author}
  {\bibfnamefont {M.}~\bibnamefont {Hochberg}}, \bibinfo {author}
  {\bibfnamefont {X.}~\bibnamefont {Sun}}, \bibinfo {author} {\bibfnamefont
  {S.}~\bibnamefont {Zhao}}, \bibinfo {author} {\bibfnamefont {H.}~\bibnamefont
  {Larochelle}}, \bibinfo {author} {\bibfnamefont {D.}~\bibnamefont
  {Englund}},\ and\ \bibinfo {author} {\bibfnamefont {M.}~\bibnamefont
  {Soljacic}},\ }\bibfield  {title} {\bibinfo {title} {Deep learning with
  coherent nanophotonic circuits},\ }\href
  {https://doi.org/10.1038/nphoton.2017.93} {\bibfield  {journal} {\bibinfo
  {journal} {Nat. Photon.}\ }\textbf {\bibinfo {volume} {11}},\ \bibinfo
  {pages} {441} (\bibinfo {year} {2017})}\BibitemShut {NoStop}%
\bibitem [{\citenamefont {Lin}\ \emph {et~al.}(2018)\citenamefont {Lin},
  \citenamefont {Rivenson}, \citenamefont {Yardimci}, \citenamefont {Veli},
  \citenamefont {Luo}, \citenamefont {Jarrahi},\ and\ \citenamefont
  {Ozcan}}]{lin2018all}%
  \BibitemOpen
  \bibfield  {author} {\bibinfo {author} {\bibfnamefont {X.}~\bibnamefont
  {Lin}}, \bibinfo {author} {\bibfnamefont {Y.}~\bibnamefont {Rivenson}},
  \bibinfo {author} {\bibfnamefont {N.~T.}\ \bibnamefont {Yardimci}}, \bibinfo
  {author} {\bibfnamefont {M.}~\bibnamefont {Veli}}, \bibinfo {author}
  {\bibfnamefont {Y.}~\bibnamefont {Luo}}, \bibinfo {author} {\bibfnamefont
  {M.}~\bibnamefont {Jarrahi}},\ and\ \bibinfo {author} {\bibfnamefont
  {A.}~\bibnamefont {Ozcan}},\ }\bibfield  {title} {\bibinfo {title}
  {All-optical machine learning using diffractive deep neural networks},\
  }\href {https://doi.org/10.1126/science.aat8084} {\bibfield  {journal}
  {\bibinfo  {journal} {Science}\ }\textbf {\bibinfo {volume} {361}},\ \bibinfo
  {pages} {1004} (\bibinfo {year} {2018})}\BibitemShut {NoStop}%
\bibitem [{\citenamefont {Colburn}\ \emph {et~al.}(2019)\citenamefont
  {Colburn}, \citenamefont {Chu}, \citenamefont {Shilzerman},\ and\
  \citenamefont {Majumdar}}]{colburn2018optical}%
  \BibitemOpen
  \bibfield  {author} {\bibinfo {author} {\bibfnamefont {S.}~\bibnamefont
  {Colburn}}, \bibinfo {author} {\bibfnamefont {Y.}~\bibnamefont {Chu}},
  \bibinfo {author} {\bibfnamefont {E.}~\bibnamefont {Shilzerman}},\ and\
  \bibinfo {author} {\bibfnamefont {A.}~\bibnamefont {Majumdar}},\ }\bibfield
  {title} {\bibinfo {title} {Optical frontend for a convolutional neural
  network},\ }\href {https://doi.org/10.1364/AO.58.003179} {\bibfield
  {journal} {\bibinfo  {journal} {Appl. Opt.}\ }\textbf {\bibinfo {volume}
  {58}},\ \bibinfo {pages} {3179} (\bibinfo {year} {2019})}\BibitemShut
  {NoStop}%
\bibitem [{\citenamefont {Kozlovsky}\ \emph {et~al.}(1988)\citenamefont
  {Kozlovsky}, \citenamefont {Nabors},\ and\ \citenamefont
  {Byer}}]{kozlovsky1988efficient}%
  \BibitemOpen
  \bibfield  {author} {\bibinfo {author} {\bibfnamefont {W.~J.}\ \bibnamefont
  {Kozlovsky}}, \bibinfo {author} {\bibfnamefont {C.}~\bibnamefont {Nabors}},\
  and\ \bibinfo {author} {\bibfnamefont {R.~L.}\ \bibnamefont {Byer}},\
  }\bibfield  {title} {\bibinfo {title} {Efficient second harmonic generation
  of a diode-laser-pumped cw nd: Yag laser using monolithic mgo: Linbo$_3$
  external resonant cavities},\ }\href {https://doi.org/10.1109/3.211}
  {\bibfield  {journal} {\bibinfo  {journal} {IEEE J. Quantum Electron}\
  }\textbf {\bibinfo {volume} {24}},\ \bibinfo {pages} {913} (\bibinfo {year}
  {1988})}\BibitemShut {NoStop}%
\bibitem [{\citenamefont {Boyd}(1992)}]{boyd1992nonlinear}%
  \BibitemOpen
  \bibfield  {author} {\bibinfo {author} {\bibfnamefont {R.~W.}\ \bibnamefont
  {Boyd}},\ }\href@noop {} {\emph {\bibinfo {title} {Nonlinear Optics}}}\
  (\bibinfo  {publisher} {Academic},\ \bibinfo {year} {1992})\BibitemShut
  {NoStop}%
\bibitem [{\citenamefont {Fryett}\ \emph {et~al.}(2016)\citenamefont {Fryett},
  \citenamefont {Seyler}, \citenamefont {Zheng}, \citenamefont {Liu},
  \citenamefont {Xu},\ and\ \citenamefont {Majumdar}}]{fryett2016silicon}%
  \BibitemOpen
  \bibfield  {author} {\bibinfo {author} {\bibfnamefont {T.~K.}\ \bibnamefont
  {Fryett}}, \bibinfo {author} {\bibfnamefont {K.~L.}\ \bibnamefont {Seyler}},
  \bibinfo {author} {\bibfnamefont {J.}~\bibnamefont {Zheng}}, \bibinfo
  {author} {\bibfnamefont {C.-H.}\ \bibnamefont {Liu}}, \bibinfo {author}
  {\bibfnamefont {X.}~\bibnamefont {Xu}},\ and\ \bibinfo {author}
  {\bibfnamefont {A.}~\bibnamefont {Majumdar}},\ }\bibfield  {title} {\bibinfo
  {title} {Silicon photonic crystal cavity enhanced second-harmonic generation
  from monolayer wse$_2$},\ }\href
  {https://doi.org/10.1088/2053-1583/4/1/015031} {\bibfield  {journal}
  {\bibinfo  {journal} {2D Materials}\ }\textbf {\bibinfo {volume} {4}},\
  \bibinfo {pages} {015031} (\bibinfo {year} {2016})}\BibitemShut {NoStop}%
\bibitem [{\citenamefont {Arnaud}(1969)}]{arnaud1969degenerate}%
  \BibitemOpen
  \bibfield  {author} {\bibinfo {author} {\bibfnamefont {J.~A.}\ \bibnamefont
  {Arnaud}},\ }\bibfield  {title} {\bibinfo {title} {Degenerate optical
  cavities},\ }\href {https://doi.org/10.1364/AO.8.000189} {\bibfield
  {journal} {\bibinfo  {journal} {Appl. Opt.}\ }\textbf {\bibinfo {volume}
  {8}},\ \bibinfo {pages} {189} (\bibinfo {year} {1969})}\BibitemShut {NoStop}%
\bibitem [{\citenamefont {Gigan}\ \emph {et~al.}(2005)\citenamefont {Gigan},
  \citenamefont {Lopez}, \citenamefont {Treps}, \citenamefont {Ma{\^\i}tre},\
  and\ \citenamefont {Fabre}}]{gigan2005image}%
  \BibitemOpen
  \bibfield  {author} {\bibinfo {author} {\bibfnamefont {S.}~\bibnamefont
  {Gigan}}, \bibinfo {author} {\bibfnamefont {L.}~\bibnamefont {Lopez}},
  \bibinfo {author} {\bibfnamefont {N.}~\bibnamefont {Treps}}, \bibinfo
  {author} {\bibfnamefont {A.}~\bibnamefont {Ma{\^\i}tre}},\ and\ \bibinfo
  {author} {\bibfnamefont {C.}~\bibnamefont {Fabre}},\ }\bibfield  {title}
  {\bibinfo {title} {Image transmission through a stable paraxial cavity},\
  }\href {https://doi.org/10.1103/PhysRevA.72.023804} {\bibfield  {journal}
  {\bibinfo  {journal} {Phys. Rev. A}\ }\textbf {\bibinfo {volume} {72}},\
  \bibinfo {pages} {023804} (\bibinfo {year} {2005})}\BibitemShut {NoStop}%
\bibitem [{\citenamefont {Chalopin}\ \emph {et~al.}(2010)\citenamefont
  {Chalopin}, \citenamefont {Chiummo}, \citenamefont {Fabre}, \citenamefont
  {Ma{\^\i}tre},\ and\ \citenamefont {Treps}}]{chalopin2010frequency}%
  \BibitemOpen
  \bibfield  {author} {\bibinfo {author} {\bibfnamefont {B.}~\bibnamefont
  {Chalopin}}, \bibinfo {author} {\bibfnamefont {A.}~\bibnamefont {Chiummo}},
  \bibinfo {author} {\bibfnamefont {C.}~\bibnamefont {Fabre}}, \bibinfo
  {author} {\bibfnamefont {A.}~\bibnamefont {Ma{\^\i}tre}},\ and\ \bibinfo
  {author} {\bibfnamefont {N.}~\bibnamefont {Treps}},\ }\bibfield  {title}
  {\bibinfo {title} {Frequency doubling of low power images using a
  self-imaging cavity},\ }\href {https://doi.org/10.1364/OE.18.008033}
  {\bibfield  {journal} {\bibinfo  {journal} {Opt. Express}\ }\textbf {\bibinfo
  {volume} {18}},\ \bibinfo {pages} {8033} (\bibinfo {year}
  {2010})}\BibitemShut {NoStop}%
\bibitem [{\citenamefont {Ghosh}\ \emph {et~al.}(2011)\citenamefont {Ghosh},
  \citenamefont {Burns}, \citenamefont {Cocker}, \citenamefont {Nimmerjahn},
  \citenamefont {Ziv}, \citenamefont {El~Gamal},\ and\ \citenamefont
  {Schnitzer}}]{ghosh2011miniaturized}%
  \BibitemOpen
  \bibfield  {author} {\bibinfo {author} {\bibfnamefont {K.~K.}\ \bibnamefont
  {Ghosh}}, \bibinfo {author} {\bibfnamefont {L.~D.}\ \bibnamefont {Burns}},
  \bibinfo {author} {\bibfnamefont {E.~D.}\ \bibnamefont {Cocker}}, \bibinfo
  {author} {\bibfnamefont {A.}~\bibnamefont {Nimmerjahn}}, \bibinfo {author}
  {\bibfnamefont {Y.}~\bibnamefont {Ziv}}, \bibinfo {author} {\bibfnamefont
  {A.}~\bibnamefont {El~Gamal}},\ and\ \bibinfo {author} {\bibfnamefont
  {M.~J.}\ \bibnamefont {Schnitzer}},\ }\bibfield  {title} {\bibinfo {title}
  {Miniaturized integration of a fluorescence microscope},\ }\href
  {https://doi.org/10.1038/nmeth.1694} {\bibfield  {journal} {\bibinfo
  {journal} {Nat. Methods}\ }\textbf {\bibinfo {volume} {8}},\ \bibinfo {pages}
  {871} (\bibinfo {year} {2011})}\BibitemShut {NoStop}%
\bibitem [{\citenamefont {Liu}\ \emph {et~al.}(2016)\citenamefont {Liu},
  \citenamefont {Sun}, \citenamefont {Majumdar},\ and\ \citenamefont
  {Sorger}}]{liu2016fundamental}%
  \BibitemOpen
  \bibfield  {author} {\bibinfo {author} {\bibfnamefont {K.}~\bibnamefont
  {Liu}}, \bibinfo {author} {\bibfnamefont {S.}~\bibnamefont {Sun}}, \bibinfo
  {author} {\bibfnamefont {A.}~\bibnamefont {Majumdar}},\ and\ \bibinfo
  {author} {\bibfnamefont {V.~J.}\ \bibnamefont {Sorger}},\ }\bibfield  {title}
  {\bibinfo {title} {Fundamental scaling laws in nanophotonics},\ }\href
  {https://doi.org/10.1038/srep37419} {\bibfield  {journal} {\bibinfo
  {journal} {Sci. Rep.}\ }\textbf {\bibinfo {volume} {6}},\ \bibinfo {pages}
  {37419} (\bibinfo {year} {2016})}\BibitemShut {NoStop}%
\bibitem [{\citenamefont {Fox}\ and\ \citenamefont
  {Li}(1961)}]{fox1961resonant}%
  \BibitemOpen
  \bibfield  {author} {\bibinfo {author} {\bibfnamefont {A.~G.}\ \bibnamefont
  {Fox}}\ and\ \bibinfo {author} {\bibfnamefont {T.}~\bibnamefont {Li}},\
  }\bibfield  {title} {\bibinfo {title} {Resonant modes in a maser
  interferometer},\ }\href {https://doi.org/10.1002/j.1538-7305.1961.tb01625.x}
  {\bibfield  {journal} {\bibinfo  {journal} {Bell Syst. Tech. J.}\ }\textbf
  {\bibinfo {volume} {40}},\ \bibinfo {pages} {453} (\bibinfo {year}
  {1961})}\BibitemShut {NoStop}%
\bibitem [{\citenamefont {Siegman}(1986)}]{siegman1986lasers}%
  \BibitemOpen
  \bibfield  {author} {\bibinfo {author} {\bibfnamefont {A.~E.}\ \bibnamefont
  {Siegman}},\ }\href@noop {} {\emph {\bibinfo {title} {Lasers}}}\ (\bibinfo
  {publisher} {University Science Books},\ \bibinfo {year} {1986})\BibitemShut
  {NoStop}%
\bibitem [{\citenamefont {Self}(1983)}]{self1983focusing}%
  \BibitemOpen
  \bibfield  {author} {\bibinfo {author} {\bibfnamefont {S.~A.}\ \bibnamefont
  {Self}},\ }\bibfield  {title} {\bibinfo {title} {Focusing of spherical
  gaussian beams},\ }\href {https://doi.org/10.1364/AO.22.000658} {\bibfield
  {journal} {\bibinfo  {journal} {Appl. Opt.}\ }\textbf {\bibinfo {volume}
  {22}},\ \bibinfo {pages} {658} (\bibinfo {year} {1983})}\BibitemShut
  {NoStop}%
\bibitem [{\citenamefont {Agrawal}\ and\ \citenamefont
  {Pattanayak}(1979)}]{agrawal1979gaussian}%
  \BibitemOpen
  \bibfield  {author} {\bibinfo {author} {\bibfnamefont {G.~P.}\ \bibnamefont
  {Agrawal}}\ and\ \bibinfo {author} {\bibfnamefont {D.~N.}\ \bibnamefont
  {Pattanayak}},\ }\bibfield  {title} {\bibinfo {title} {Gaussian beam
  propagation beyond the paraxial approximation},\ }\href
  {https://doi.org/10.1364/JOSA.69.000575} {\bibfield  {journal} {\bibinfo
  {journal} {J. Opt. Soc. Am.}\ }\textbf {\bibinfo {volume} {69}},\ \bibinfo
  {pages} {575} (\bibinfo {year} {1979})}\BibitemShut {NoStop}%
\bibitem [{\citenamefont {Phillips}\ and\ \citenamefont
  {Eliasson}(2018)}]{phillips2018camera}%
  \BibitemOpen
  \bibfield  {author} {\bibinfo {author} {\bibfnamefont {J.~B.}\ \bibnamefont
  {Phillips}}\ and\ \bibinfo {author} {\bibfnamefont {H.}~\bibnamefont
  {Eliasson}},\ }\href@noop {} {\emph {\bibinfo {title} {Camera Image Quality
  Benchmarking}}}\ (\bibinfo  {publisher} {John Wiley \& Sons},\ \bibinfo
  {year} {2018})\BibitemShut {NoStop}%
\bibitem [{\citenamefont {Nozaki}\ \emph {et~al.}(2010)\citenamefont {Nozaki},
  \citenamefont {Tanabe}, \citenamefont {Shinya}, \citenamefont {Matsuo},
  \citenamefont {Sato}, \citenamefont {Taniyama},\ and\ \citenamefont
  {Notomi}}]{nozaki2010sub}%
  \BibitemOpen
  \bibfield  {author} {\bibinfo {author} {\bibfnamefont {K.}~\bibnamefont
  {Nozaki}}, \bibinfo {author} {\bibfnamefont {T.}~\bibnamefont {Tanabe}},
  \bibinfo {author} {\bibfnamefont {A.}~\bibnamefont {Shinya}}, \bibinfo
  {author} {\bibfnamefont {S.}~\bibnamefont {Matsuo}}, \bibinfo {author}
  {\bibfnamefont {T.}~\bibnamefont {Sato}}, \bibinfo {author} {\bibfnamefont
  {H.}~\bibnamefont {Taniyama}},\ and\ \bibinfo {author} {\bibfnamefont
  {M.}~\bibnamefont {Notomi}},\ }\bibfield  {title} {\bibinfo {title}
  {Sub-femtojoule all-optical switching using a photonic-crystal nanocavity},\
  }\href {https://doi.org/10.1038/nphoton.2010.89} {\bibfield  {journal}
  {\bibinfo  {journal} {Nat. Photon.}\ }\textbf {\bibinfo {volume} {4}},\
  \bibinfo {pages} {477} (\bibinfo {year} {2010})}\BibitemShut {NoStop}%
\bibitem [{\citenamefont {Majumdar}\ and\ \citenamefont
  {Rundquist}(2014)}]{majumdar2014cavity}%
  \BibitemOpen
  \bibfield  {author} {\bibinfo {author} {\bibfnamefont {A.}~\bibnamefont
  {Majumdar}}\ and\ \bibinfo {author} {\bibfnamefont {A.}~\bibnamefont
  {Rundquist}},\ }\bibfield  {title} {\bibinfo {title} {Cavity-enabled
  self-electro-optic bistability in silicon photonics},\ }\href
  {https://doi.org/10.1364/OL.39.003864} {\bibfield  {journal} {\bibinfo
  {journal} {Opt. Lett.}\ }\textbf {\bibinfo {volume} {39}},\ \bibinfo {pages}
  {3864} (\bibinfo {year} {2014})}\BibitemShut {NoStop}%
\bibitem [{\citenamefont {Fryett}\ \emph {et~al.}(2015)\citenamefont {Fryett},
  \citenamefont {Dodson},\ and\ \citenamefont {Majumdar}}]{fryett2015cavity}%
  \BibitemOpen
  \bibfield  {author} {\bibinfo {author} {\bibfnamefont {T.~K.}\ \bibnamefont
  {Fryett}}, \bibinfo {author} {\bibfnamefont {C.~M.}\ \bibnamefont {Dodson}},\
  and\ \bibinfo {author} {\bibfnamefont {A.}~\bibnamefont {Majumdar}},\
  }\bibfield  {title} {\bibinfo {title} {Cavity enhanced nonlinear optics for
  few photon optical bistability},\ }\href
  {https://doi.org/10.1364/OE.23.016246} {\bibfield  {journal} {\bibinfo
  {journal} {Opt. Express}\ }\textbf {\bibinfo {volume} {23}},\ \bibinfo
  {pages} {16246} (\bibinfo {year} {2015})}\BibitemShut {NoStop}%
\bibitem [{\citenamefont {Liu}\ \emph {et~al.}(2019)\citenamefont {Liu},
  \citenamefont {Zheng}, \citenamefont {Chen}, \citenamefont {Fryett},\ and\
  \citenamefont {Majumdar}}]{liu2019van}%
  \BibitemOpen
  \bibfield  {author} {\bibinfo {author} {\bibfnamefont {C.-h.}\ \bibnamefont
  {Liu}}, \bibinfo {author} {\bibfnamefont {J.}~\bibnamefont {Zheng}}, \bibinfo
  {author} {\bibfnamefont {Y.}~\bibnamefont {Chen}}, \bibinfo {author}
  {\bibfnamefont {T.}~\bibnamefont {Fryett}},\ and\ \bibinfo {author}
  {\bibfnamefont {A.}~\bibnamefont {Majumdar}},\ }\bibfield  {title} {\bibinfo
  {title} {Van der waals materials integrated nanophotonic devices},\ }\href
  {https://doi.org/10.1364/OME.9.000384} {\bibfield  {journal} {\bibinfo
  {journal} {Opt. Mater. Express}\ }\textbf {\bibinfo {volume} {9}},\ \bibinfo
  {pages} {384} (\bibinfo {year} {2019})}\BibitemShut {NoStop}%
\bibitem [{\citenamefont {Jaksch}\ \emph {et~al.}(1998)\citenamefont {Jaksch},
  \citenamefont {Bruder}, \citenamefont {Cirac}, \citenamefont {Gardiner},\
  and\ \citenamefont {Zoller}}]{jaksch1998cold}%
  \BibitemOpen
  \bibfield  {author} {\bibinfo {author} {\bibfnamefont {D.}~\bibnamefont
  {Jaksch}}, \bibinfo {author} {\bibfnamefont {C.}~\bibnamefont {Bruder}},
  \bibinfo {author} {\bibfnamefont {J.~I.}\ \bibnamefont {Cirac}}, \bibinfo
  {author} {\bibfnamefont {C.~W.}\ \bibnamefont {Gardiner}},\ and\ \bibinfo
  {author} {\bibfnamefont {P.}~\bibnamefont {Zoller}},\ }\bibfield  {title}
  {\bibinfo {title} {Cold bosonic atoms in optical lattices},\ }\href
  {https://doi.org/10.1103/PhysRevLett.81.3108} {\bibfield  {journal} {\bibinfo
   {journal} {Phys. Rev. Lett.}\ }\textbf {\bibinfo {volume} {81}},\ \bibinfo
  {pages} {3108} (\bibinfo {year} {1998})}\BibitemShut {NoStop}%
\bibitem [{\citenamefont {Jia}\ \emph {et~al.}(2018)\citenamefont {Jia},
  \citenamefont {Schine}, \citenamefont {Georgakopoulos}, \citenamefont {Ryou},
  \citenamefont {Clark}, \citenamefont {Sommer},\ and\ \citenamefont
  {Simon}}]{jia2018strongly}%
  \BibitemOpen
  \bibfield  {author} {\bibinfo {author} {\bibfnamefont {N.}~\bibnamefont
  {Jia}}, \bibinfo {author} {\bibfnamefont {N.}~\bibnamefont {Schine}},
  \bibinfo {author} {\bibfnamefont {A.}~\bibnamefont {Georgakopoulos}},
  \bibinfo {author} {\bibfnamefont {A.}~\bibnamefont {Ryou}}, \bibinfo {author}
  {\bibfnamefont {L.~W.}\ \bibnamefont {Clark}}, \bibinfo {author}
  {\bibfnamefont {A.}~\bibnamefont {Sommer}},\ and\ \bibinfo {author}
  {\bibfnamefont {J.}~\bibnamefont {Simon}},\ }\bibfield  {title} {\bibinfo
  {title} {A strongly interacting polaritonic quantum dot},\ }\href
  {https://doi.org/10.1038/s41567-018-0071-6} {\bibfield  {journal} {\bibinfo
  {journal} {Nat. Phys.}\ }\textbf {\bibinfo {volume} {14}},\ \bibinfo {pages}
  {550} (\bibinfo {year} {2018})}\BibitemShut {NoStop}%
\bibitem [{\citenamefont {Yu}\ and\ \citenamefont
  {Capasso}(2014)}]{yu2014flat}%
  \BibitemOpen
  \bibfield  {author} {\bibinfo {author} {\bibfnamefont {N.}~\bibnamefont
  {Yu}}\ and\ \bibinfo {author} {\bibfnamefont {F.}~\bibnamefont {Capasso}},\
  }\bibfield  {title} {\bibinfo {title} {Flat optics with designer
  metasurfaces},\ }\href {https://doi.org/10.1038/nmat3839} {\bibfield
  {journal} {\bibinfo  {journal} {Nat. Mater.}\ }\textbf {\bibinfo {volume}
  {13}},\ \bibinfo {pages} {139} (\bibinfo {year} {2014})}\BibitemShut
  {NoStop}%
\bibitem [{\citenamefont {Zhan}\ \emph {et~al.}(2017)\citenamefont {Zhan},
  \citenamefont {Colburn}, \citenamefont {Dodson},\ and\ \citenamefont
  {Majumdar}}]{zhan2017metasurface}%
  \BibitemOpen
  \bibfield  {author} {\bibinfo {author} {\bibfnamefont {A.}~\bibnamefont
  {Zhan}}, \bibinfo {author} {\bibfnamefont {S.}~\bibnamefont {Colburn}},
  \bibinfo {author} {\bibfnamefont {C.~M.}\ \bibnamefont {Dodson}},\ and\
  \bibinfo {author} {\bibfnamefont {A.}~\bibnamefont {Majumdar}},\ }\bibfield
  {title} {\bibinfo {title} {Metasurface freeform nanophotonics},\ }\href
  {https://doi.org/10.1038/s41598-017-01908-9} {\bibfield  {journal} {\bibinfo
  {journal} {Sci. Rep.}\ }\textbf {\bibinfo {volume} {7}},\ \bibinfo {pages}
  {1673} (\bibinfo {year} {2017})}\BibitemShut {NoStop}%
\bibitem [{\citenamefont {Colburn}\ \emph {et~al.}(2018)\citenamefont
  {Colburn}, \citenamefont {Zhan},\ and\ \citenamefont
  {Majumdar}}]{colburn2018metasurface}%
  \BibitemOpen
  \bibfield  {author} {\bibinfo {author} {\bibfnamefont {S.}~\bibnamefont
  {Colburn}}, \bibinfo {author} {\bibfnamefont {A.}~\bibnamefont {Zhan}},\ and\
  \bibinfo {author} {\bibfnamefont {A.}~\bibnamefont {Majumdar}},\ }\bibfield
  {title} {\bibinfo {title} {Metasurface optics for full-color computational
  imaging},\ }\href {https://doi.org/10.1126/sciadv.aar2114} {\bibfield
  {journal} {\bibinfo  {journal} {Sci. Adv.}\ }\textbf {\bibinfo {volume}
  {4}},\ \bibinfo {pages} {eaar2114} (\bibinfo {year} {2018})}\BibitemShut
  {NoStop}%
\bibitem [{\citenamefont {Goodman}(2008)}]{goodman2008introduction}%
  \BibitemOpen
  \bibfield  {author} {\bibinfo {author} {\bibfnamefont {J.}~\bibnamefont
  {Goodman}},\ }\href@noop {} {\emph {\bibinfo {title} {Introduction to Fourier
  Optics}}}\ (\bibinfo  {publisher} {McGraw Hill},\ \bibinfo {year}
  {2008})\BibitemShut {NoStop}%
\bibitem [{\citenamefont {Sziklas}\ and\ \citenamefont
  {Siegman}(1975)}]{sziklas1975mode}%
  \BibitemOpen
  \bibfield  {author} {\bibinfo {author} {\bibfnamefont {E.~A.}\ \bibnamefont
  {Sziklas}}\ and\ \bibinfo {author} {\bibfnamefont {A.}~\bibnamefont
  {Siegman}},\ }\bibfield  {title} {\bibinfo {title} {Mode calculations in
  unstable resonators with flowing saturable gain. 2: Fast fourier transform
  method},\ }\href {https://doi.org/10.1364/AO.14.001874} {\bibfield  {journal}
  {\bibinfo  {journal} {Appl. Opt.}\ }\textbf {\bibinfo {volume} {14}},\
  \bibinfo {pages} {1874} (\bibinfo {year} {1975})}\BibitemShut {NoStop}%
\bibitem [{\citenamefont {Matsushima}\ and\ \citenamefont
  {Shimobaba}(2009)}]{matsushima2009band}%
  \BibitemOpen
  \bibfield  {author} {\bibinfo {author} {\bibfnamefont {K.}~\bibnamefont
  {Matsushima}}\ and\ \bibinfo {author} {\bibfnamefont {T.}~\bibnamefont
  {Shimobaba}},\ }\bibfield  {title} {\bibinfo {title} {Band-limited angular
  spectrum method for numerical simulation of free-space propagation in far and
  near fields},\ }\href {https://doi.org/10.1364/OE.17.019662} {\bibfield
  {journal} {\bibinfo  {journal} {Opt. Express}\ }\textbf {\bibinfo {volume}
  {17}},\ \bibinfo {pages} {19662} (\bibinfo {year} {2009})}\BibitemShut
  {NoStop}%
\end{thebibliography}%


\appendix

\section{Cavity simulation details}

\subsection{Light propagation}

The simulated optical field is given by a square grid of complex numbers, where the dimension size is determined by the aperture diameter $a$ divided by the grid resolution. The grid solution is set to $\lambda/2 = 500$ $\mu$m.

The propagation of a field from one plane to another is carried out using the angular spectrum approach, which is based on the fact that any field $u(x,y)$ in real space can be represented as a sum of plane waves with different $k$ vectors \cite{goodman2008introduction}:
\begin{equation}
u(x,y) = \iint U(k_x, k_y) e^{i(k_x x + k_y y)} \, dx \, dy
\end{equation}

Each plane wave propagates over a distance $z$, which can be modeled by multiplying the field by the propagator $H = e^{ik_z z}$, where
\begin{equation}
k_z = \sqrt{k^2 - k_x^2 - k_z^2}
\end{equation}
and $k = 2\pi/\lambda$. Thus, to propagate a field, we take the Fourier transform of the field, multiply it by $H$, and take the inverse Fourier transform.

\subsection{Optical elements}

Each optical element of the cavity, shown in Fig. 1a, can be represented as a phase mask. The phase profile of the flat mirror is unity. The phase profile of the lens is given by
\begin{equation}
\phi_{\text{lens}} (f) = - \frac{\pi \left(x^2 + y^2\right)}{f\lambda }
\end{equation}
where $x$ and $y$ denote the transverse coordinates in the simulation grid. The negative sign ensures that the lens is converging. The phase profile of the curved mirror is given by that of a lens with $f = R/2$. 

Besides the phase profile, the optical elements also have reflectivities that simulate optical loss from either reflection or transmission. For simplicity, the reflectivities for both the flat and the curved mirrors are set equal to $r$. For the lens, the reflectivity is set to zero so that all light is transmitted without loss.

Thus, the action of an optical element on an incident field is to multiply it by the phase profile and the reflectivity pixel-by-pixel:
\begin{equation}
u'(x,y) = r e^{i\phi(x,y)} u(x,y)
\end{equation}

Additionally, we implement a sharp-edged circular aperture, whose diameter $a$ is equal to the side length of the square grid, by incorporating a circular mask on the plane of each optical element.

\begin{figure}
\includegraphics[width=83mm]{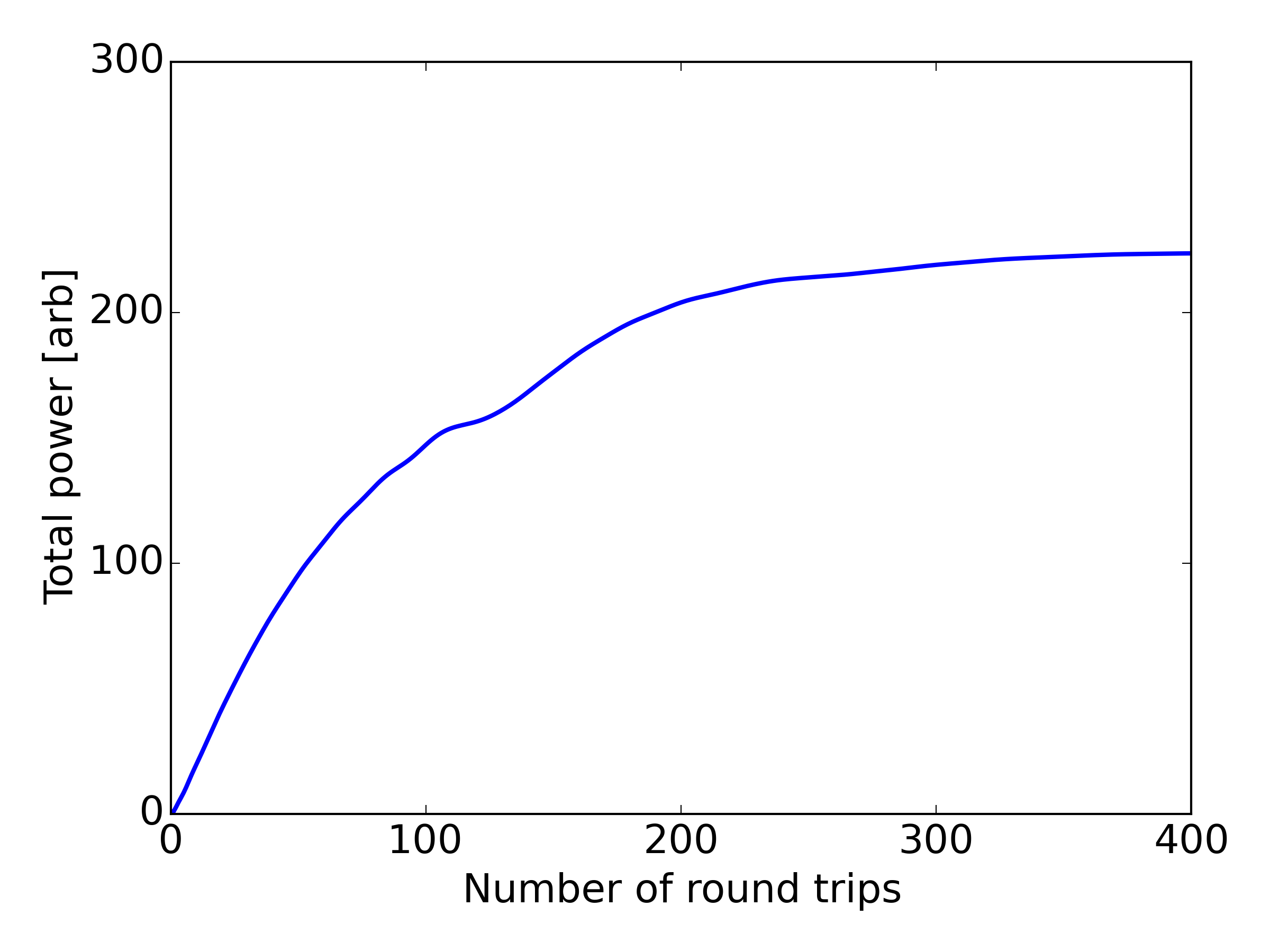}
\caption{\label{Figure:cavity_cartoon} (Color online) \textbf{Intracavity power buildup}. Plot of the intracavity power (sum over all intensity values in the simulation grid at the plane of the flat mirror) vs. the number of cavity round trips. The power increases rapidly and saturates after a sufficient number of round trips, determined by the cavity finesse, or, in terms of the simulation variables, the amplitude reflection coefficient $r$ of the two cavity mirrors.} 
\end{figure}


\subsection{Cavity round trip}

A simulation begins with an incident field $u_{inc}$ that has unity value everywhere in the two-dimensional grid. The field enters the cavity through the flat mirror, with the transmission amplitude coefficient given by $t = \sqrt{1-r^2}$. The transmitted wave is then ready to make the first round trip inside the cavity.

The round trip consists of a series of alternating free-space propagations and actions by the optical elements. The sequence is as follows: (1) propagation by distance $d_1$ from the flat mirror to the lens; (2) action of the lens; (3) propagation by distance $d_2$ from the lens to the curved mirror; (4) action of the curved mirror; (5) propagation by distance $d_2$ from the curved mirror to the lens; (6) action of the lens; (7) propagation by distance $d_1$ from the lens to the flat mirror; and finally (8) action of the flat mirror. Together the eight steps comprise one round trip around the cavity. 

The total field at the plane of the flat mirror is the sum of the individual fields after making successive round trips. The total intensity is the absolute square of the total field. The total power can be calculated by adding up all the intensity values of the pixels in the grid. We end the simulation when the total power approaches a constant value that indicates that a steady state has been reached between enhancement and loss; see Fig. 6 for the rise and saturation of the total power for a typical cavity mode.


\section{Paraxial degenerate cavity}

There are several types of transverse mode degeneracies, depending on the number of modes that share the same Gouy phase. According to paraxial cavity theory, the mode frequencies of a cavity are given by 
\begin{equation}
v_{qmn} = \left( q + (m+n+1) \frac{\alpha}{2\pi} \right) v_{FSR}
\end{equation}
where $q$ denotes the axial mode, and $m$ and $n$ ($l$ and $p$ in the case of LG modes) denote the transverse mode. The free spectral range is given by $v_{FSR} = c/L$, where $L$ is the cavity's round trip distance. The Gouy phase $\alpha$ is related to the eigenvalues of the cavity's round-trip $ABCD$ matrix via:
\begin{equation}
\alpha = \cos^{-1} \left( \frac{A+D}{2} \right)
\end{equation}
where $A$ and $D$ are the matrix's diagonal elements \cite{siegman1986lasers}.

Hence, different degeneracies can be realized by designing a cavity such that $\alpha = 2\pi \, K/N$, where $K$ and $N$ are integers \cite{gigan2005image}. For our cavity shown in Fig. 1a, the ABCD matrix becomes a 2 x 2 unit matrix when $f = R = d_1/2 = d_2/2$, which leads to $\alpha=0$. Such a cavity is said to be self-imaging, since all its transverse modes are completely degenerate, and in principle, any image can be transmitted, and amplified, without distortion. In the geometric optics picture, an arbitrary ray of light in a self-imaging cavity re-traces and returns to the same displacement and slope upon one round trip.





\begin{figure}
\includegraphics[width=83mm]{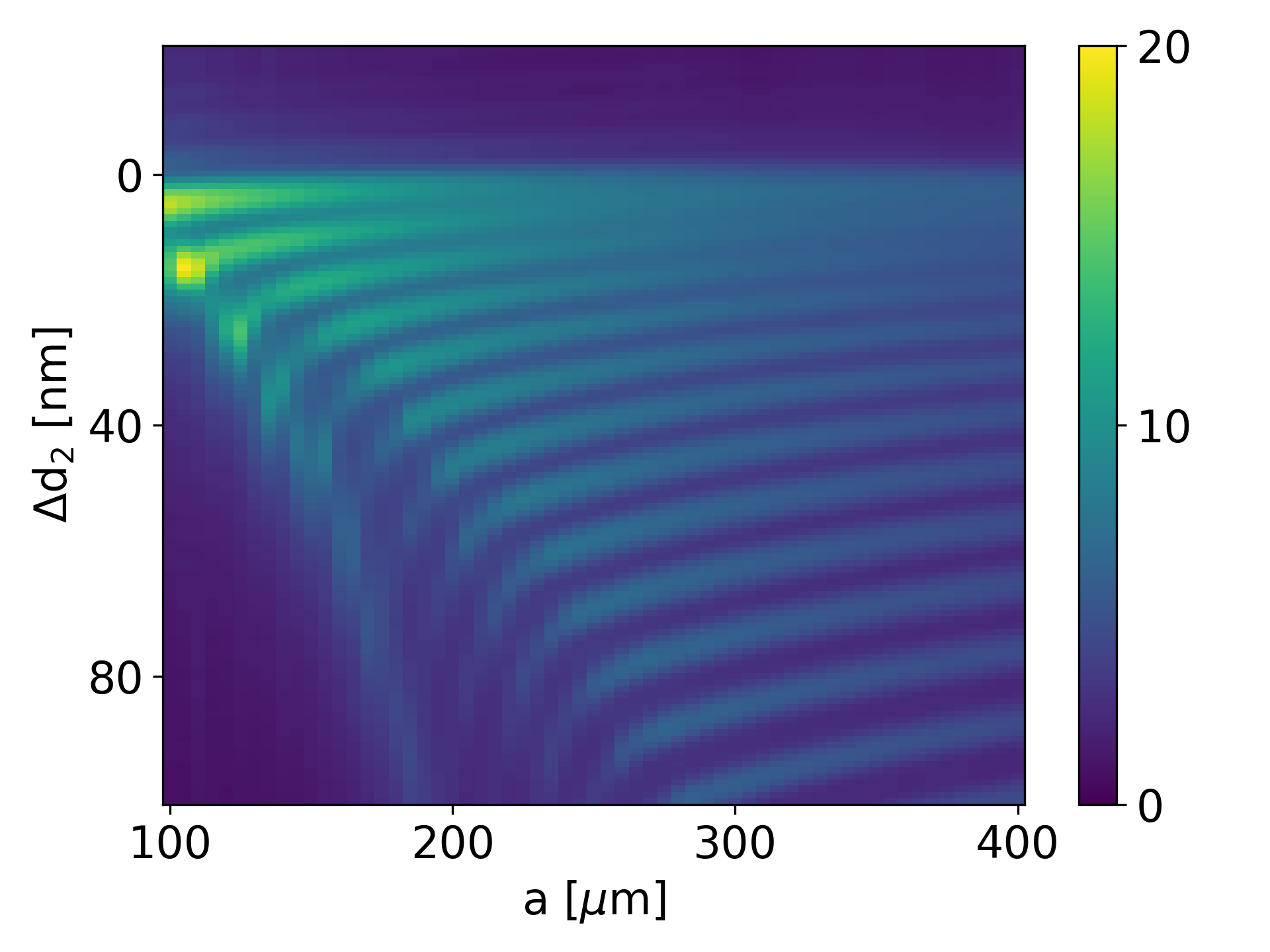}
\caption{\label{Figure:cavity_cartoon} (Color online) \textbf{Mode spectrum vs. aperture size}. Plot of the cavity mode spectrum as a function of the detuning $\Delta d_2$ (y-axis) and the aperture diameter (x-axis). As the aperture diameter increases, more peaks are observed in the spectrum, indicating the onset of progressively higher-order modes. At the same time, the modes move closer to one another, even when the size of the aperture starts to become more than an order-of-magnitude bigger than the beam width. The colorbar represents the square root of the field intensity.} 
\end{figure}

\section{Finite aperture effect}

For practical laser cavities with macroscopic mirrors, the aperture can be safely assumed to be infinite. On the other hand, the simulated cavity is necessarily finite in transverse extent due to the limitation in computational resources. Interestingly, the finite aperture results in a numerical effect that appears as a contribution to the spread in the modes. Figure 7 shows the intensity spectrum of a cavity at the predicted degeneracy condition ($f = R = d_1/2 = 1$ mm, $r = 0.99$, $d_2 = d_1 + \Delta d_2$) as a function of $\Delta d_2$ (y-axis) and the aperture diameter $a$ (x-axis). As $a$ increases, more modes can ``fit'' inside the cavity and begin to resonate. In addition, the modes continue to shift and move closer to one another, even when the aperture is an order-of-magnitude larger than the modes.

This effect remains even when we modify the angular spectrum approach with zero-padding the simulation grid, to linearize the discrete Fourier transform \cite{sziklas1975mode}, and imposing a limit on the field bandwidth \cite{matsushima2009band}. Both methods have been developed to prevent stray light from leaking into ``neighboring'' cells of the simulation. Further studies are necessary to understand why the size of the simulation grid affects the mode locations, but as can be seen in Fig. 7, the numerical effect is small: a few-nanometer shift in the modes when $a$ changes from 200 $\mu$m to 400 $\mu$m ) vs. a tens-of-nanometers shift in the modes for a few-micron tuning of $d_2$ in Fig. 2.







\begin{figure}
\includegraphics[width=83mm]{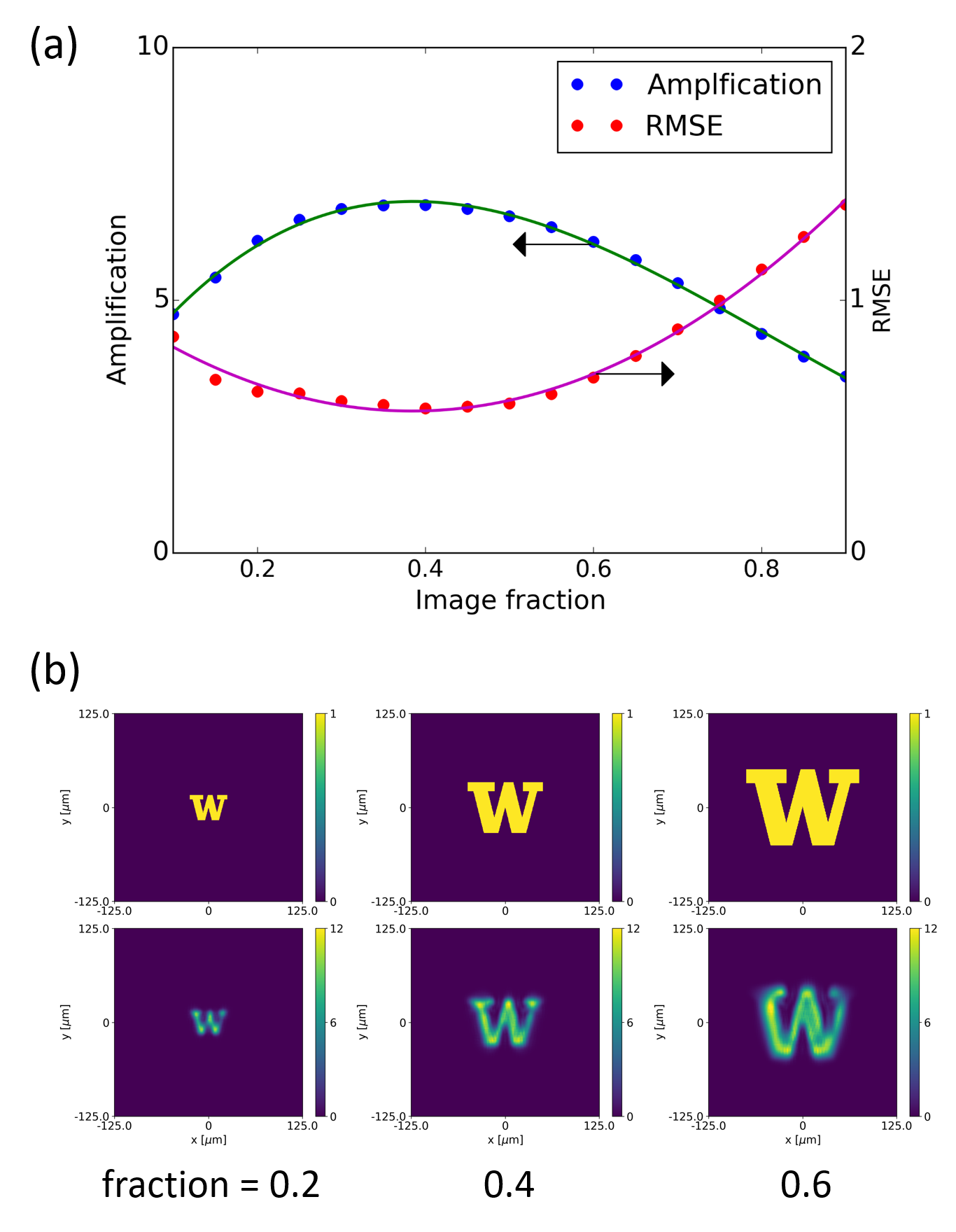}
\caption{\label{Figure:cavity_cartoon} (Color online) \textbf{Intracavity image quality vs incident image size}. (a) Plot of the amplification (blue) and the RMSE (red) of the intracavity image as a function of the size of the incident image, as its lateral size is changed from 0.1 to 0.9 of the aperture diameter. The error bars are smaller than the points. The curves are guides to the eyes only. (b) Transverse profiles of the incident image and the intracavity field intensity for different incident image sizes. All three profiles exhibit similar amplification, but the profile for the fractional size of 0.4 exhibits the highest fidelity. The colorbars indicate the intensity.} 
\end{figure}

\section{Input image}

The image chosen for the simulations is a stylized version of the letter ``W''. 

As stated in the main text, the amplification and the fidelity of the intracavity image are heavily dependent on the size of the cavity, which determines the spread in the cavity's transverse modes. They are also dependent on the modal composition of the incident image itself. The more modes that go into the image's make up, the greater the distortion in the intracavity image. For the simulations in the main text, the size of the incident image has been fixed to be a third of the grid size. Here we further explore the effect of the size of the incident image on the intracavity image.   


Figure 8a shows the amplification (blue) and the RMSE (red) of the intracavity field versus the fractional size of the incident image relative to the aperture diameter $a$. A fraction of $0.2$, for instance, indicates that the x-dimension of the incident image is $20\%$ of the x-dimension of $a = 250$ $\mu$m. All the other parameters remain fixed during the simulation: $f = R = d_1/2 = 2$ mm and $r = 0.95$. As expected, the amplification and the RMSE have opposite trends. When the incident image is too small (fraction $\sim 0.2$), it excites many cavity modes, and as a result, the amplification is low and the RMSE is high. As the fraction increases to about 0.4, the size of the incident image becomes comparable to the beam width of the lowest cavity modes. Thus, it excites only a few modes, yielding the highest amplification and the lowest RMSE. After hitting this sweet spot, further increasing the incident image size reverses the trends, yielding low amplification and high RMSE again. Figure 8b shows the transverse profiles of the incident image and intracavity field intensities for three different incident image sizes.


\end{document}